\documentclass[%
superscriptaddress,
 preprint,
showkeys,showpacs,preprintnumbers,
amsmath,amssymb,
aps,
prb,
]{revtex4-2}

\usepackage{graphicx}% Include figure files
\usepackage{dcolumn}% Align table columns on decimal point
\usepackage{bm}% bold math
\usepackage{upgreek}
\usepackage{amsmath}

\begin{document}

%\preprint{APS/123-QED}

\title{High-efficiency infrared upconversion imaging with nonlinear silicon metasurfaces empowered by quasi-bound states in the continuum}% Force line breaks with \\

\author{Tingting Liu}
%\email{ttliu@ncu.edu.cn}
\affiliation{School of Information Engineering, Nanchang University, Nanchang 330031, China}
\affiliation{Institute for Advanced Study, Nanchang University, Nanchang 330031, China}

\author{Jumin Qiu}
\email{qiujumin@email.ncu.edu.cn}
\affiliation{School of Physics and Materials Science, Nanchang University, Nanchang 330031, China}

\author{Meibao Qin}
\affiliation{School of Education, Nanchang Institute of Science and Technology, Nanchang 330108, China}
\affiliation{School of Physics and Materials Science, Nanchang University, Nanchang 330031, China}

\author{Xu Tu}
\affiliation{Institute for Advanced Study, Nanchang University, Nanchang 330031, China}

\author{Huifu Qiu}
\affiliation{Institute for Advanced Study, Nanchang University, Nanchang 330031, China}

\author{Feng Wu}
\affiliation{School of Optoelectronic Engineering, Guangdong Polytechnic Normal University, Guangzhou 510665, China}

\author{Tianbao Yu}
\affiliation{School of Physics and Materials Science, Nanchang University, Nanchang 330031, China}

\author{Qiegen Liu}
\email{liuqiegen@ncu.edu.cn}
\affiliation{School of Information Engineering, Nanchang University, Nanchang 330031, China}

\author{Shuyuan Xiao}
\email{syxiao@ncu.edu.cn}
\affiliation{School of Information Engineering, Nanchang University, Nanchang 330031, China}
\affiliation{Institute for Advanced Study, Nanchang University, Nanchang 330031, China}

\begin{abstract}

Infrared imaging is indispensable for its ability to penetrate obscurants and visualize thermal signatures, yet its practical use is hindered by the intrinsic limitations of conventional detectors. Nonlinear upconversion, which converts infrared light into the visible band, offers a promising pathway to address these challenges. Here, we demonstrate high-efficiency infrared upconversion imaging using nonlinear silicon metasurfaces. By strategically breaking in-plane symmetry, the metasurface supports a high-$Q$ quasi-bound states in the continuum resonance, leading to strongly enhanced third-harmonic generation (THG) with a conversion efficiency of $3\times10^{-5}$ at a pump intensity of 10 GW/cm$^{2}$. Through this THG process, the metasurface enables high-fidelity upconversion of arbitrary infrared images into the visible range, achieving a spatial resolution of $\sim 6$ $\upmu$m as verified using a resolution target and various customized patterns. This work establishes a robust platform for efficient nonlinear conversion and imaging, highlighting the potential of CMOS-compatible silicon metasurfaces for high-performance infrared sensing applications with reduced system complexity.

\end{abstract}

%\pacs{42.70.-a, 42.79.-e, 78.67.Pt}% PACS, the Physics and Astronomy
%Use showkeys class option if keyword
\maketitle

%\tableofcontents

\section{\label{sec1}Introduction}

Infrared (IR) imaging holds significant application value across diverse fields such as night vision, thermal inspection, biomedical diagnostics, and remote sensing, owing to its ability to penetrate certain obscurants and directly visualize thermal signatures. However, the inherent limitations of conventional IR detectors, particularly their low sensitivity, high cost, and the need for cryogenic cooling in some spectral bands, pose substantial challenges for practical implementation. Nonlinear frequency upconversion imaging, which effectively transforms IR light into the visible spectrum where highly sensitive and mature silicon-based detectors operate, represents a crucial technological pathway\cite{Barh2019}. To circumvent the limitations associated with stringent phase matching in traditional bulk nonlinear crystals, nonlinear imaging utilizing nanostructured ultrathin metasurfaces has garnered considerable research interest. Metasurfaces excel in precise and versatile wavefront manipulation of light at the nanoscale, enabling unprecedented control over amplitude\cite{Zheng2021, Deng2024, Liu2024}, phase\cite{Deng2018, Li2024}, polarization\cite{Ding2020, Wang2023, Li2025a}, and orbital angular momentum\cite{Ren2020, Meng2025}. Initial research efforts in nonlinear imaging metasurfaces primarily focused on harnessing this wavefront engineering capability. By strategically encoding spatial patterns into the nanostructures to locally control the amplitude, phase, or polarization of the light fields, researchers successfully demonstrated various nonlinear optical functionalities, including nonlinear holography and image encoding\cite{Gao2018, Xu2019, Mao2022}, metalenses\cite{Schlickriede2020, Tseng2022}, edge detection\cite{Zhou2022}, and even asymmetric imaging\cite{Kruk2022, Rocco2024}. While these pioneering works validate the concept of metasurface-enabled nonlinear wavefront shaping for IR imaging, they often face limitations such as relatively low conversion efficiency, limited number of predesigned images encoded in a single metasurface, complex design and fabrication requirements for multi-dimensional control.

In contrast to the wavefront-shaping approach, direct upconversion imaging based on nonlinear metasurfaces is currently emerging as a promising alternative strategy. This paradigm leverages resonant metasurfaces to boost the nonlinear conversion efficiency via significantly enhance the local electromagnetic field, thereby reducing the requirement for high incident pump power\cite{Vabishchevich2018, Liu2019, Koshelev2019, Huang2022, Qu2022, Liu2023}. Crucially, this approach circumvents the need for complex spatial encoding of amplitude, phase, or polarization across the metasurface to form specific image displays. Instead, the metasurface efficiently converts any incident infrared image pattern pixel-by-pixel into its visible counterpart. This intrinsic parallelism simplifies design and relaxes fabrication tolerances, effectively mitigating challenges associated with structural non-uniformities that plague complex wavefront-encoded designs. Camacho-Morales et al. explored the feasibility of this concept using Gallium Arsenide (GaAs) metasurfaces for IR upconversion imaging via the coherent parametric process of sum-frequency generation (SFG)\cite{CamachoMorales2021}. However, the performance of such resonant metasurfaces is fundamentally constrained by the limited quality ($Q$) factors of their localized Mie resonances and the inherent material absorption of GaAs within the visible spectrum. Enabling efficient upconversion imaging via metasurfaces with high-$Q$ resonances promises substantial advantages due to the strong nonlinearity enhancement. In the framework, novel concepts leveraging high-$Q$ physics, such as quasi-bound states in the continuum (quasi-BICs) and nonlocal resonances, have been rapidly introduced into the field\cite{CamachoMorales2022, SantiagoCruz2022, Zhang2022a, Huang2023a, Feng2023, Sanderson2024, ValenciaMolina2024, Sun2025}. Implementations on silicon (Si) platforms are particularly attractive due to its CMOS compatibility, low visible-wavelength absorption, and well-established mature fabrication processes. For instance, Zheng et al. demonstrated enhanced third-harmonic generation (THG) and four-wave mixing (FWM) nonlinear conversion and imaging using quasi-BIC resonances in silicon metasurfaces\cite{Zheng2023, Zheng2024}. Despite these significant advances, the absolute conversion efficiencies reported to date are typically on the order of 10$^{-6}$ or lower, indicating considerable room for improvement toward meeting the requirements of practical applications.

In this work, we propose and experimentally demonstrate high-efficiency IR upconversion imaging with high-$Q$ silicon metasurfaces. Leveraging the quasi-BIC resonance derived from a symmetry-protected BIC combined with the strong third-order optical nonlinearity of silicon, we achieve enhanced THG to covert near-IR light to the visible wavelength range. We provide a comprehensive analysis of the formation mechanism of the BIC, and by precisely breaking the structural symmetry along only one direction—unlike conventional designs that break in-plane symmetry in both directions—we significantly reduce radiative loss and achieve quasi-BIC resonances with much higher $Q$-factors. We experimentally obtain an absolute conversion efficiency of THG of $3\times10^{-5}$ under a pump intensity of 10 GW/cm$^{2}$ from the designed metasurface. The larger mode volume arising from strong electromagnetic field confinement within the silicon nanostructures provides a powerful platform for nonlinear imaging. Specifically, when employing an IR pump beam modulated with the image of a target at the quasi-BIC wavelength, the spatial information of the target is efficiently upconverted via THG into the visible range. This process yields high-efficiency imaging with excellent image quality. Our results establish a new approach for developing efficient nonlinear silicon metasurfaces, and highlight their significant potential of nonlinear nanophotonics towards advanced IR imaging techniques.

\section{\label{sec2}Results and discussion}

\subsection{\label{sec2.1}Design principle of quasi-BIC metasurfaces}

Figure 1(a) schematically illustrates the proposed nonlinear metasurface for IR upconversion imaging. Infrared light carrying the spatial information of a target object is incident on the metasurface, generating a third-harmonic generation (THG) signal that upconverts the image features into the visible range. Under ideal conditions without wavefront distortion during conversion, the upconverted image can be directly captured using a standard visible camera. The inset shows the energy level diagram of the THG process—a third-order nonlinear optical effect that triples the photon energy. We adopt silicon as the nonlinear material due to its strong third-order nonlinear susceptibility ($\chi^{(3)}$), which has been widely exploited in processes such as THG and four-wave mixing\cite{Xu2018, Xu2022, Liu2023a, Moretti2024}. The metasurface unit cell, depicted in Fig. 1(b), comprises a pair of circular-elliptical nanodisks with a height of 220 nm, arranged periodically on a glass substrate. The array periods along the $x$ and $y$ directions are $P_{x} = 1000$ nm and $P_{y} = 500$ nm, respectively. The radius of the circular nanodisk and the major axis of the elliptical nanodisk are both set to $R = 180$ nm. Structural asymmetry is introduced solely along one direction by varying the minor axis $r$ of the elliptical nanodisk along $x$. This symmetry breaking establishes a controlled radiative channel, converting the genuine BIC into a high-$Q$ quasi-BIC resonance. %Compared to earlier designs breaking symmetry along both directions, this unidirectional approach reduces radiative loss and significantly increases the $Q$-factor, thereby enhancing the light-matter interaction in both linear and nonlinear regimes. 

\begin{figure*}[htbp]
	\centering
	\includegraphics[width=\linewidth]{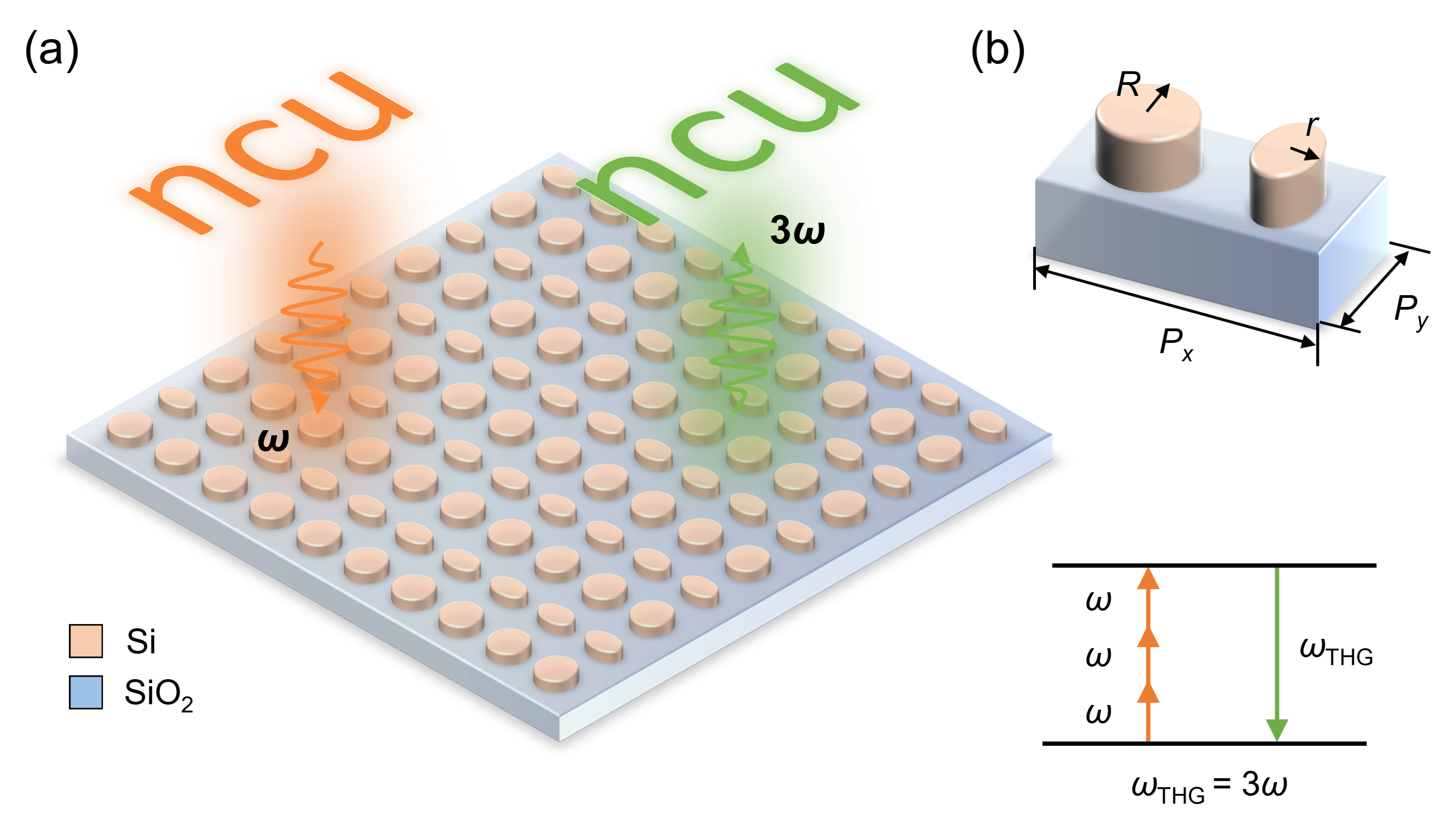}
	\caption{Quasi-BIC metasurfaces for IR to visible upconversion imaging. (a) The schematic illustration of the THG imaging based on the as-designed metasurfaces. Inset: the energy level diagram of the THG process, where $\omega$ and $\omega_{\text{THG}}$ are the angular frequencies of the incident and the nonlinear generated photons, respectively. (b) The unit cell of the metasurface consisting of a periodic array of silicon circular-elliptical nanodisks placed on top of a glass substrate.}
	\label{fig1}
\end{figure*}

To validate the design, we begin with a theoretical analysis to elucidate the formation of the quasi-BIC resonance in the metasurface. Numerical methods are detailed in the Supporting Information. Figure 2(a) shows the evolution of the resonant modes as the minor axis $r$ is varied, with the error bars indicating the radiative loss (i.e., the inverse radiation lifetime of the mode). Variations in $r$ shift the eigenfrequency and modify the radiation loss of the resonant system. When $r = R = 180$ nm, the metasurface supports a genuine BIC that fully decouples from free space, corresponding to a non-radiative mode. This BIC is symmetry-protected: as shown in Fig. 2(b), its magnetic field distribution $H_{z}$ exhibits even symmetry under 180$^{\circ}$ rotation around the $z$-axis. In contrast, the electromagnetic fields of a plane wave propagating along $z$—the only available radiation channel at the $\Gamma$ point in the subdiffractive regime—are odd under the same rotation. Due to this fundamental symmetry mismatch, the BIC mode cannot couple to a normally incident plane wave of any linear polarization\cite{He2018, Overvig2020}. Introducing in-plane asymmetry by varying the minor axis $r$ along the $x$-direction converts the genuine BIC into a leaky quasi-BIC resonance, opening a radiative channel. The radiation loss increases gradually with the deviation of $r$ from $R$, which is further corroborated by the corresponding magnetic field distribution $H_{z}$, where the in-plane inversion symmetry $(x, y) \rightarrow (-x, -y)$ of $H_{z}$ at the BIC condition is broken by the structural perturbation. For comparison, we also analyze the case with symmetry breaking along both directions using a metasurface composed of a pair of circular nanodisks with different radii (refer to the Supporting Information). By confining structural asymmetry to only one direction, radiative loss is further suppressed, leading to stronger light–matter interactions in both linear and nonlinear regimes.

\begin{figure*}[htbp]
	\centering
	\includegraphics[width=\linewidth]{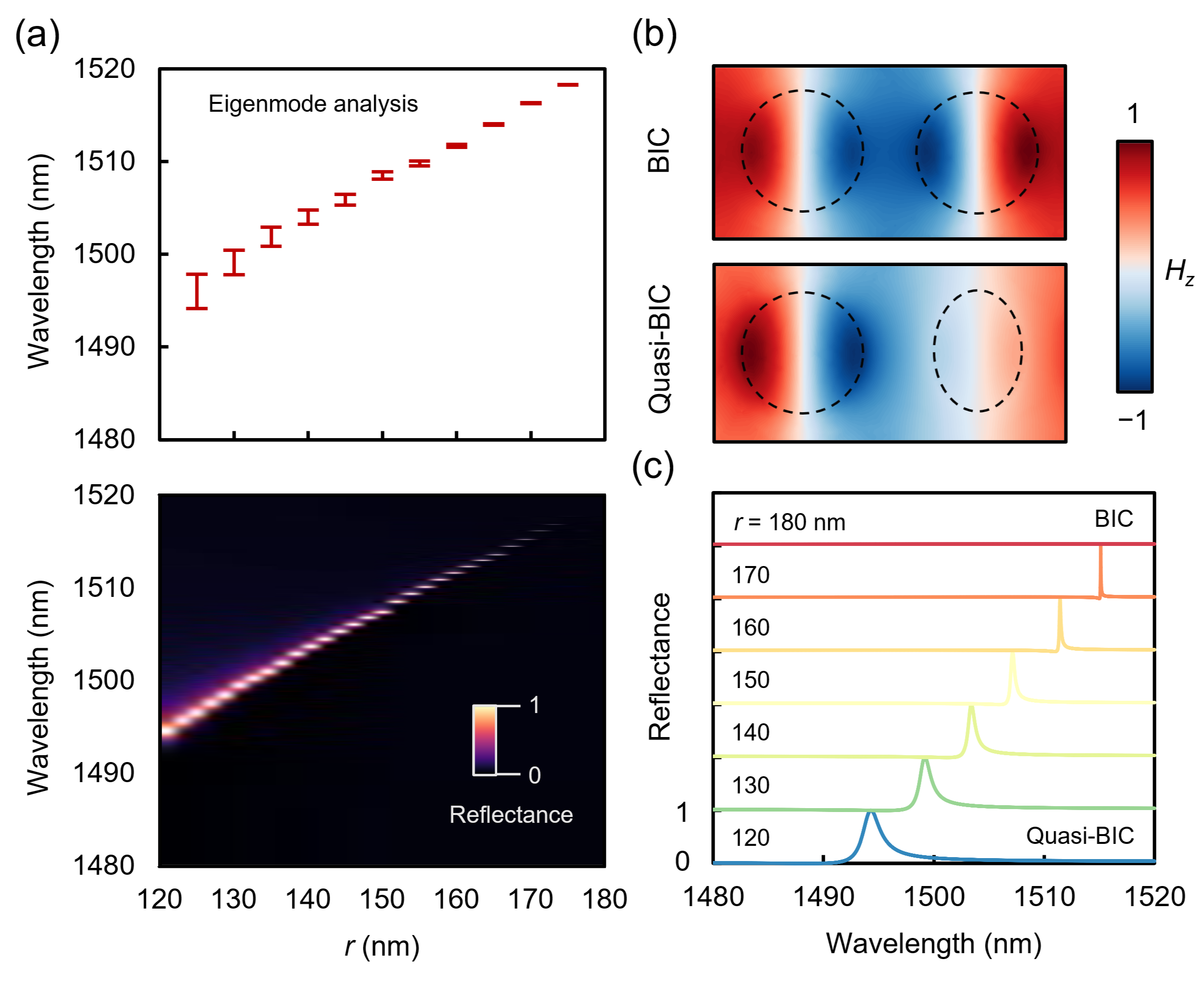}
	\caption{Mode analysis of the quasi-BIC metasurfaces. (a) The eigenfrequency and linear reflectance spectra as a function of the minor axis $r$ of the elliptical nanodisk. The error bars represent the relative magnitude of the mode radiation loss during $r$ variations. (b) The distribution of magnetic fields in the $z$-direction $H_{z}$ for the genuine BIC at $r = 180$ nm and quasi-BIC at $r = 120$ nm, respectively. (c) The evolution of reflectance spectra as $r$ deviates from $R$ from 180 nm to 120 nm.}
	\label{fig2}
\end{figure*}

Guided by the eigenmode analysis, we perform numerical simulations of the linear optical responses of the metasurfaces under a normally incident $y$-polarized plane wave. As shown in Fig. 2(a), when the metasurface is symmetric with a radius of $r = R = 180$ nm, the reflectance exhibits a vanishing peak with a zero linewidth, corresponding to a genuine BIC that remains completely decoupled from external radiation. As the minor axis $r$ gradually decreases, introducing structural asymmetry, the radiation loss increases. This leads to a broadening of the resonant linewidth and a concurrent blueshift in the resonance wavelength. Figure 2(c) details the spectral evolution, showing how the reflectance transforms as $r$ deviates from $R$ from 180 nm to 120 nm in increments of 10 nm. It explicitly confirms the transformation of the ideal BIC into a finite-$Q$ quasi-BIC resonance, which is characterized by a distinct asymmetric Fano lineshape. This universal behavior, consistent with the eigenmode analysis, demonstrates that the $Q$-factor of the quasi-BIC systematically decreases with increasing asymmetry in the proposed metasurfaces.

\subsection{\label{sec2.2}Fabrication and linear reflectance measurement}

Building on the numerical design, we conduct experimental demonstrations of the quasi-BIC metasurfaces. The fabrication of metasurface samples follows the nanofabrication process flow in Fig. 3(a). It starts from a silicon-on-insulator (SOI) substrate, which features a 220 nm top silicon layer, a 2 $\upmu$m buried oxide layer, and a silicon handle wafer approximately 700 $\upmu$m in thickness. After spin-coating a layer of ZEP520 photoresist onto the cleaned SOI substrate, electron-beam lithography (EBL) is used to pattern periodic nanostructures into the photoresist. These patterns are then transferred into the top silicon layer via inductively coupled plasma (ICP) etching, where the photoresist acts as the etch mask. Finally, any residual resist is stripped by immersing the samples in N-methyl-2-pyrrolidone (NMP). A set of metasurface samples with the minor axis $r$ ranging from 120 nm to 180 nm, in increments of 10 nm, are fabricated, each with a size of $800 \times 800$ $\upmu$m$^{2}$. The scanning electron microscope (SEM) images of representative metasurface samples with $r = 180$ nm (symmetric case) and $r = 120$ nm (asymmetric case) are shown in Fig. 3(b), demonstrating high uniformity and minimal surface roughness.

\begin{figure*}[htbp]
	\centering
	\includegraphics[width=\linewidth]{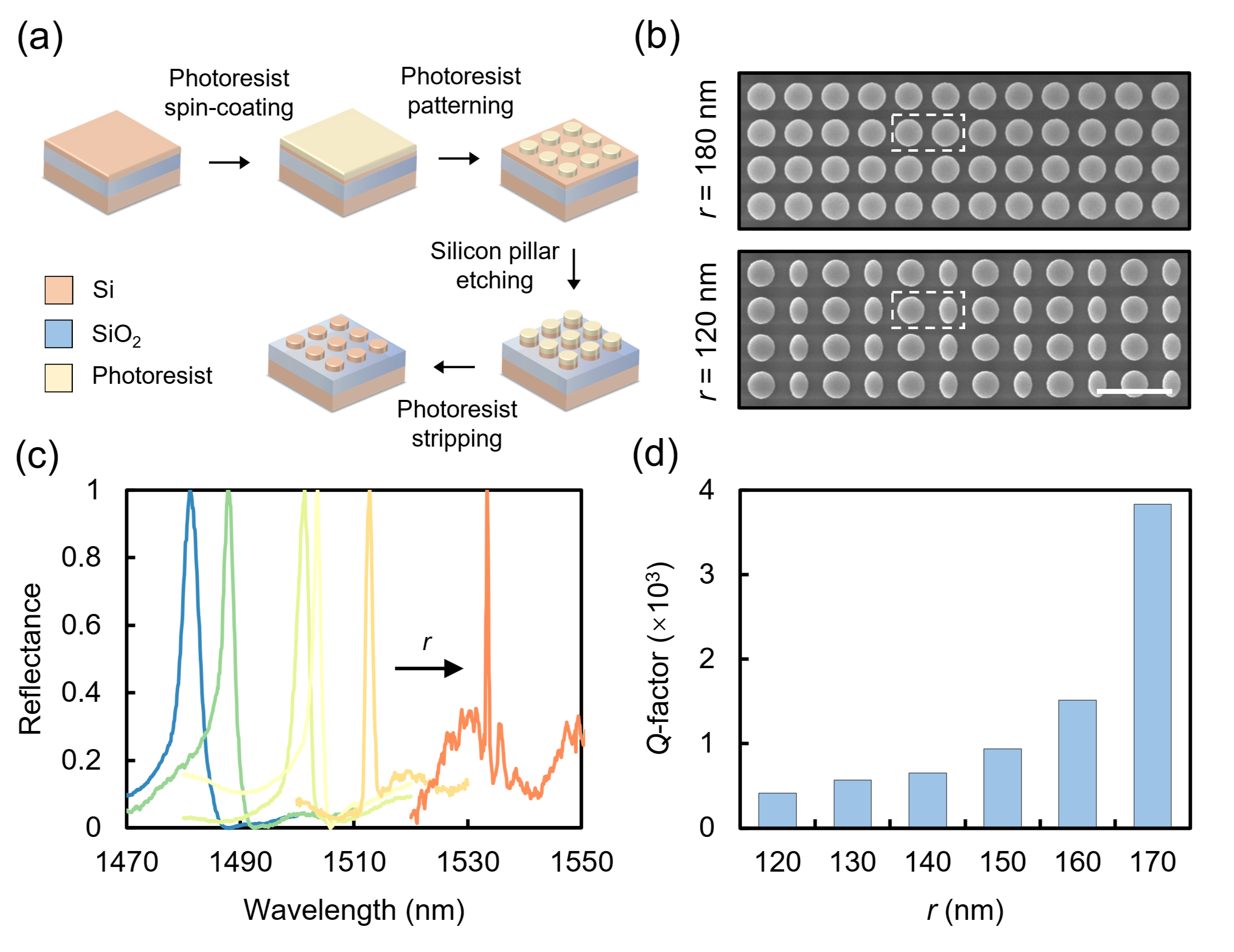}
	\caption{Fabrication and measured linear optical responses of the quasi-BIC metasurfaces. (a) The schematic of the nanofabrication process flow on an SOI wafer. (b) The SEM images of the fabricated metasurface samples in the top view with the minor axis $r = 180$ nm (symmetric case) and $r = 120$ nm (asymmetric case). Scale bar: 1 $\upmu$m. (c) The reflectance spectra of the metasurfaces with $r$ ranging from 120 nm to 170 nm (increasing $r$ from left to right) under $y$-polarized incidence. (d) The $Q$-factors extracted from the spectra in (c) at different $r$.}
	\label{fig3}
\end{figure*}

We measured the linear reflectance spectra of the fabricated metasurfaces using a custom-built optical setup with a broadband light source linearly-polarized along the $y$ direction (see details in the Supporting Information). Fig. 3(c) shows the measured reflectance spectra with the minor axis $r$ ranges from 120 nm to 170 nm in increments of 10 nm. By fitting the spectra with the Fano formula (refer to the Supporting Information), the corresponding $Q$-factors are extracted, as summarized in Fig. 3(d). It can be observed that as $r$ approaches $R$, the linewidth of the quasi-BIC resonances narrows dramatically, accompanied by higher $Q$-factors and stronger electric field confinement. This behavior is consistent with the simulated results in Fig. 2 and further confirms the symmetry-protected BIC origin of the resonances\cite{Hsu2016, Huang2023}. Notably, the measured $Q$-factor reaches a high value of up to $\sim 4000$ for $r = 170$ nm and remains around $\sim 500$ even under the highest structural asymmetry with $r = 120$ nm. The relatively high $Q$-factor originates from suppressed radiation loss, which results from precisely breaking the structural symmetry along only one direction—a strategy that limits energy leakage channels and is highly advantageous for nonlinear conversion and imaging applications.

\subsection{\label{sec2.3}Nonlinear THG measurement}

Next we proceed to the nonlinear optical responses of the same set of metasurface samples. The THG measurement is conducted in the reflection mode. A femtosecond fiber laser coupled to an optical parametric amplifier, which provides high intensity and ultrashort pulse duration, is used as the light source to pump the THG process (see details in the Supporting Information). The central wavelength of the pump beam can be tuned to align with the quasi-BIC resonance wavelength of each sample across a range from 1470 nm to 1550 nm, with a repetition rate of $\sim 200$ kHz and a pulse duration of $\sim 200$ fs within the near-infrared region of interest. The normalized conversion efficiency, defined as $\zeta=P_{\text{THG}}^{\text{peak}}/(P_{\text{Pump}}^{\text{peak}})^{3}$, represents the ratio between the output peak power at the THG wavelength and incident peak power at the pump wavelength. This metric is independent of the pump power and emphasizes the intrinsic role of field enhancement within the metasurfaces in driving the nonlinear THG process\cite{Liu2019, Carletti2019}. In Fig. 4(a), $\zeta$ steadily increases as $r$ approaches $R$, a trend consistent with the corresponding increase in the $Q$-factor derived from linear reflectance measurement in Fig. 3(d). These results confirm that the enhanced THG process originates from the strongly confined electromagnetic fields supported by the high-$Q$ quasi-BIC resonances. Notably, an intriguing and somewhat counterintuitive phenomenon is observed. Although the $Q$-factor increases sharply as $r$ approaches $R$—consistent with the universal scaling law of quasi-BICs in symmetry-broken metasurfaces—the growth of normalized conversion efficiency gradually slows under the same condition. This discrepancy can be attributed to a spectral mismatch between the narrowing resonance bandwidth and the finite spectral width of the laser pulses. As the $Q$-factor rises, the resonance linewidth becomes significantly narrower than the pump bandwidth, leading to incomplete coupling of the pulse energy into the resonant mode. This effect underscores the practical limitation imposed by the available laser source on the nonlinear conversion performance\cite{Anthur2020, Liu2021}.

\begin{figure*}[htbp]
	\centering
	\includegraphics[width=\linewidth]{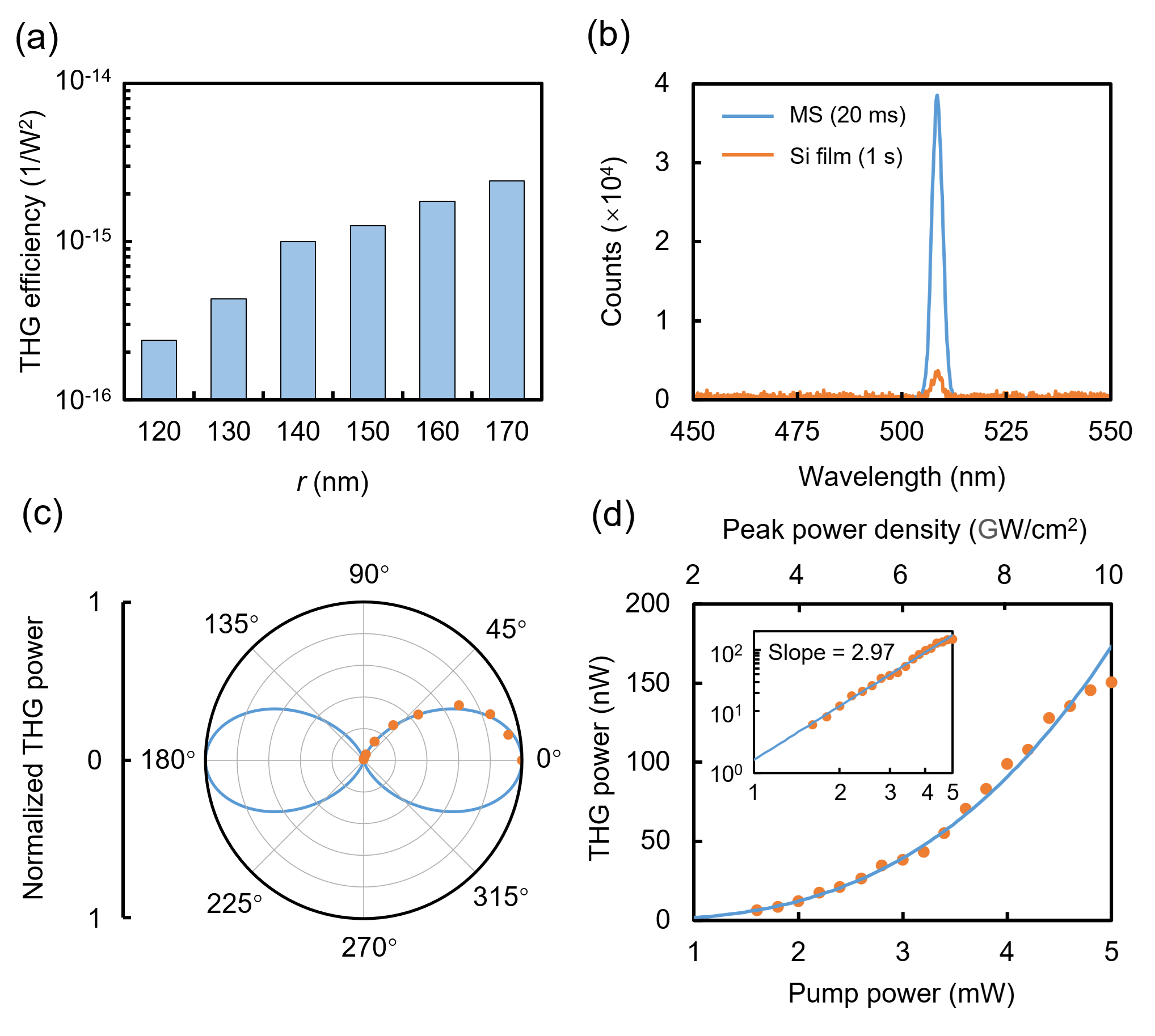}
	\caption{Measured nonlinear optical responses of the quasi-BIC metasurfaces. (a) The normalized THG conversion efficiency of the metasurfaces with the minor axis $r$ ranges from 120 nm to 170 nm under $y$-polarized incidence. (b) The THG spectra of the metasurface with $r = 170$ nm and the unpatterned silicon film. (c) The normalized THG power profile under different polarization angles with $r = 170$ nm. (d) The THG power as a function of pump power with $r = 170$ nm. Inset shows power dependence of the THG process in logarithmic scale.}
	\label{fig4}
\end{figure*}

Figure 4(b) presents a comparison between the measured THG spectrum of the metasurface sample with $r = 170$ nm and that of an unpatterned silicon film of equivalent thickness. Both are excited by a pump beam centered at 1533 nm, matching the quasi-BIC resonance wavelength of the metasurface. While a peak appears at 511 nm in both spectra, the THG signal intensity from the quasi-BIC metasurface is over 650 times greater than that from the flat film, further validating the substantial role of the localized field arising from the quasi-BIC metasurfaces in enhancing nonlinear optical effects. We also quantify the dependence of this resonantly enhanced THG signal on the polarization of the pump beam. As shown in Fig. 4(c), the THG power exhibits a typical letter '8' pattern when the pump polarization is rotated, with the maximum intensity obtained under $y$-polarized incidence. This behavior arises from the polarization-sensitive nature of the designed quasi-BIC metasurfaces, which are most effectively excited by $y$-polarized light, as indicated in the mode analysis in Fig. 2. 

In Fig. 4(d), we examine the power dependence of THG for this metasurface. The THG power increases substantially with the pump power. By fitting the experimental data to the function $P_{\text{THG}}^{\text{average}}=a(P_{\text{Pump}}^{\text{average}})^{b}$, we obtain an exponent $b=2.97$, indicating an approximately cubic relationship consistent with the theoretical scaling law for the third-order nonlinear process. This cubic dependence is further supported by the inset logarithmic plot, in which the fitted slope of 2.97 confirms the THG nature of the process. The pump power varies about 1 mW to 5 mW, and the corresponding pump intensity can be estimated based on the pulse duration, repetition rate, and spot size of the pump beam. At an average pump power of 5 mW, the average collected THG power reaches 150 nW. Under these conditions, the conversion efficiency, defined as $\eta=P_{\text{THG}}^{\text{average}}/P_{\text{Pump}}^{\text{average}}$, is calculated to be $3\times10^{-5}$ at a pump intensity of approximately 10 GW/cm². This performance is comparable to state-of-the-art silicon-based nonlinear metasurfaces\cite{Tang2024, Wang2024, Deng2024a}, and represents a threefold improvement over the efficiency previously reported for silicon nanodisk arrays in our work\cite{Liu2025}. It is noted that the THG power shows a tendency toward saturation at the highest pump power level. This phenomenon, which has also been observed in previous studies, leads to a gradual deviation from the ideal cubic scaling law\cite{Hail2024}. Nonlinear absorption mechanisms, primarily two-photon absorption (TPA) and free-carrier absorption (FCA), are identified as the main causes of this saturation.

\subsection{\label{sec2.4}Nonlinear imaging}

Following the demonstration of the enhanced THG process by the quasi-BIC metasurfaces, we finally evaluate the nonlinear upconversion imaging performance. The experimental setup is schematically illustrated in Fig. 5(a). We use the metasurface sample with the minor axis $r = 170$ nm, under $y$-polarization incidence tuned to 1533 nm for maximizing the nonlinear conversion efficiency. Using a lens group, the incident pump power can be distributed across a large area on the metasurface with an expanded beam spot. The infrared image of the test targest incident on the metasurface is upconverted to the visible range via the THG process and subsequently captured by an sCMOS camera. Further details of the custom-built optical setup are provided in the Supporting Information. A Siemens star resolution target is first employed as a representative object. As shown in Fig. 5(b), this target contains radially converging lines whose spatial frequency increases toward the center, making it well-suited for assessing imaging resolution. When the metasurface overlaps with this target, the pattern in the overlapping area is converted to a visible image with clear edges. Here a green color map is used to represent the peak wavelength of the resonantly enhanced THG signal at 511 nm. The upconverted visible image clearly resolves fine stripes, yielding a spatial resolution on the order of 6 $\upmu$m. It is noteworthy that the stripe width at the metasurface plane covers roughly a finite array of $6\times12$ unit cells, where the boundary effect is deemed insignificant on the field enhancement and thus the conversion efficiency by the quasi-BIC resonance. Furthermore, due to the polarization-sensitive nature of the designed quasi-BIC metasurface, a clear image is obtained under $y$-polarized incidence, while no signal is detected under $x$-polarized incidence. In addition, the nonlinear imaging is observed only from the metasurface region, in sharp contrast with the dark background where the unpatterned silicon layer yields no detectable signal. 

\begin{figure*}[htbp]
	\centering
	\includegraphics[width=\linewidth]{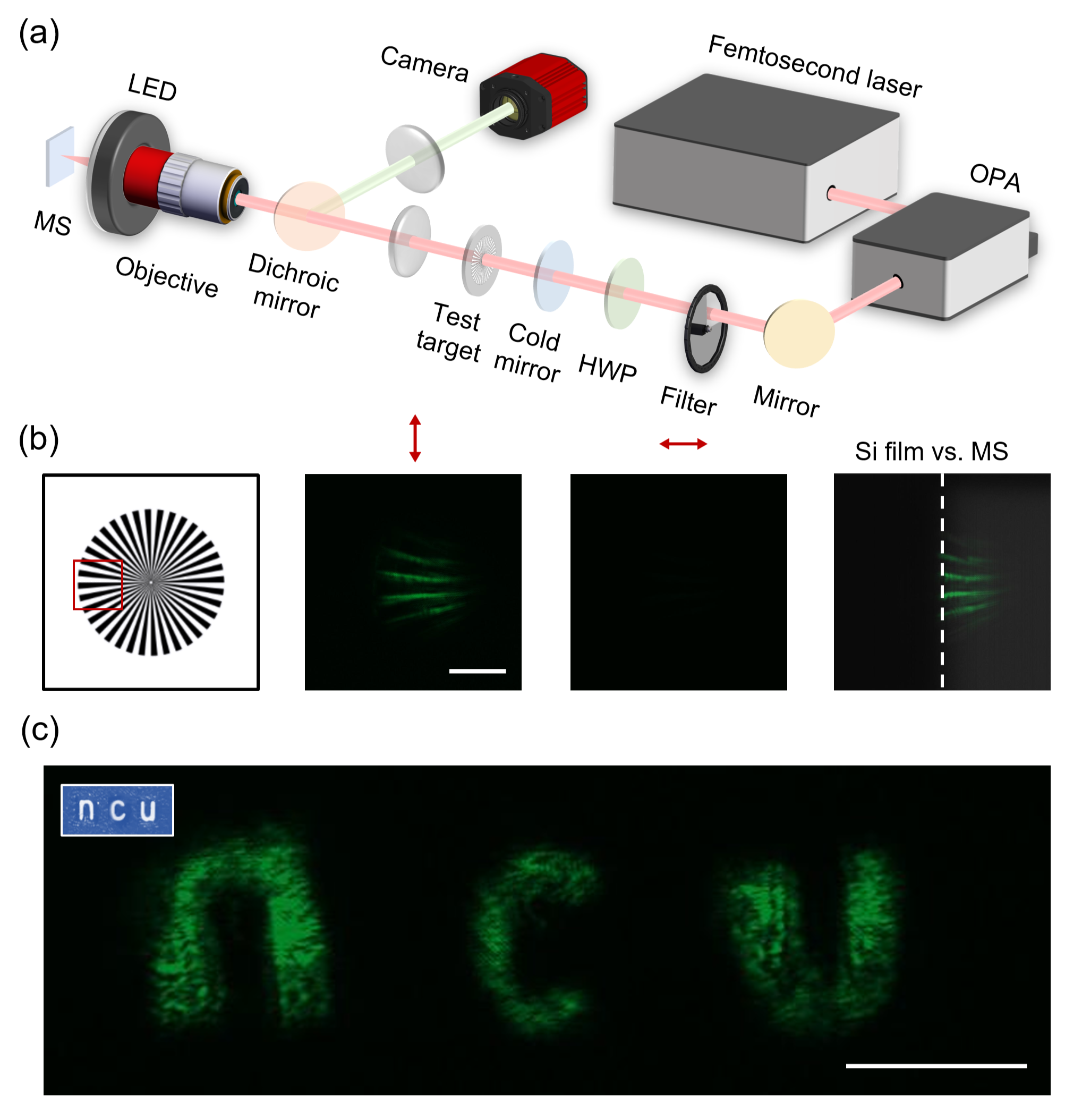}
	\caption{Measured IR upconversion imaging with the quasi-BIC metasurfaces. (a) The schematic of the custom-built optical setup for nonlinear imaging. (b) The Siemens star resolution target and its corresponding upconverted visible images under the following three cases: the metasurface under $y$-polarized incidence, the metasurface under $x$-polarized incidence, and a comparison between an unpatterned silicon film and the metasurface (both under $y$-polarized incidence). (c) The upconverted visible image of letters 'NCU', and the 3D-printed target in the inset. Scale bar: 100 $\upmu$m.}
	\label{fig5}
\end{figure*}

The versatility of this nonlinear imaging strategy is further demonstrated by upconverting an arbitrary object. As shown in Fig. 5(c), the letters 'NCU' (abbreviation for Nanchang University) are clearly resolved. These results confirm the capability of the quasi-BIC metasurface to convert arbitrary IR images into the visible range via a resonantly enhanced THG process. This technique enables direct detection of IR light using silicon-based visible cameras. The high conversion efficiency results from the high-$Q$ quasi-BIC resonance and strong field confinement, which collectively support high-resolution nonlinear imaging. Notably, our metasurfaces are fabricated using a mature SOI process, facilitating practical applications. Moreover, unlike SFG imaging involved with the pump light amplifying the signal light, our THG approach requires only a single pump beam, significantly reducing system complexity. These advantages underscore the potential of this platform in infrared sensing and imaging applications.

\section{\label{sec3}Conclusions}

In conclusion, we have demonstrated high-efficiency infrared upconversion imaging using CMOS-compatible silicon metasurfaces that support quasi-bound states in the continuum. By strategically breaking in-plane symmetry along only one direction—unlike conventional bidirectional symmetry-breaking designs—we achieve high-$Q$ quasi-BIC resonances that significantly suppress radiative loss and strongly enhance local field confinement. This approach leads to a remarkable THG efficiency of $3\times10^{-5}$ at 10 GW/cm$^{2}$. Furthermore, the metasurface facilitates pixel-by-pixel upconversion of arbitrary infrared images into the visible range via THG, achieving a spatial resolution of $\sim 6$ $\upmu$m, as validated using a standard resolution target and various customized patterns. This work not only establishes a new paradigm for designing high-efficiency nonlinear metasurfaces but also highlights a scalable and practical platform for infrared imaging with substantially reduced system complexity. Note that the principle of precisely-controlled symmetry breaking can be broadly extended to enhance other nonlinear processes, such as second-harmonic and high-harmonic generations\cite{Yang2019, Zograf2022, Zalogina2023, Jangid2024}, and can be further augmented through integration with two-dimensional materials\cite{Bernhardt2020, Lochner2021}, van der Waals materials\cite{Fan2025, Wang2025}, or electro-optical materials\cite{He2024, Zhang2025, DiFrancescantonio2025, Qiu2025}, to enable dynamically reconfigurable nonlinear imaging.

\begin{acknowledgments}	
	
This work was supported by the National Natural Science Foundation of China (Grants No. 12304420, No. 12264028, No. 12364045, No. 12364049, and No. 12104105), the Natural Science Foundation of Jiangxi Province (Grants No. 20232BAB201040, No. 20232BAB211025, and No. 20242BAB25041), and the Young Elite Scientists Sponsorship Program by JXAST (Grants No. 2023QT11 and No. 2025QT04). 

The authors wish to extend their profound gratitude to Professor Lei Xu at Nottingham Trent University. His invaluable guidance and meticulous attention to detail in the design and execution of the nonlinear imaging experiments were instrumental to this work.

\end{acknowledgments}

%\bibliography{Ref}% Produces the bibliography via BibTeX.


\end{document}